# Design of a SOIMUMPs Inertial Sensor and readout Charge Amplifier

Helder F. Campos, Nuno M. Paulino and João F. Loureiro
helder.campos@fe.up.pt, nmcp@fe.up.pt
Faculty of Engineering of the University of Porto (FEUP)

*Abstract*—This paper presents the design and post-layout characteristics of a differential capacitance based inertial accelerometer This includes a MEMS based mechanical sensing element and a CMOS charge amplifier, which is the first stage of a readout circuit. The mechanical sensor is designed according to the SOIMUMPs fabrication process technology, and the readout circuit targeted AMS 0.35μm technology. Post layout simulations indicated a ±5G dynamic range, a maximum bandwidth of 1.58 kHz, non-linearity of 0.077% and a resolution of 10.5 μG/ √ Hz. The readout circuit charge amplifier is fully differential and incorporated in a switched capacitor (SC) topology with CDS.

## I. INTRODUCTION

INERTIAL accelerometer sensors are currently widely used, and are essential components for many applications, from vibration monitoring for predictive maintenance and navigation guidance to entertainment systems, amongst others [1]. Due to widespread deployment of these devices, low manufacture costs are a requirement, while simultaneously maintaining acceptable performance characteristics for end applications. Thus, many fabrication processes, sensor technologies and readout topologies have been developed in order to attain competitive characteristics at reduced cost [2].

Microelectromechanical systems (MEMS) are the technological choice for implementation of not only acceleration sensors but also pressure, flow or other types of sensors. These devices can be designed in surface or bulk micromachining technologies [3]-[8]. Surface micromachining is based on standard deposition of layers of materials. Specifically in this context, this allows for design of higher proof masses in smaller areas, and higher resolution systems. Its multilayer characteristic also provides more alternatives for integration of sensors, actuator and other mechanisms in the same wafer occupying smaller areas, thus reducing manufacture costs. Unlike surface micromachining, bulk micromachining comprises a relatively simpler fabrication process and a lower cost associated with design and fabrication [2].

Similarly, different kinds of sensor transduction mechanisms exist. And can be grouped as mechanical (piezoresistive/piezoelectric properties of materials, capacitance variation sensing or resonance sensing), thermal (measure of thermal property variations due to movement of heated air) or magnetic (magnetoresitivity). [1-4].

For inertial accelerometers specifically, piezoresistivity and differential capacitor schemes are prominent. Capacitive accelerometers are very efficient for high precision and low noise applications due to their high sensitivity, low-temperature sensitivity and simple structure. These transduction mechanisms cause minute changes in the mechanical system, which then need to be coupled to a readout chain that amplifies, filters and converts this information into a digital output, if desired. Joint fabrication of CMOS devices and MEMS devices in the same wafer is possible, facilitating this integration, although fabrication in separate wafers is also possible, followed by a wafer bonding step. [9].

For the closed loop amplification state, opting for a capacitor based feedback as opposed to a resistor scheme introduces several advantages. First, designs that employ both resistors and capacitors as part of the signal sampling path are subject to nominal value variations due to fabrication. Since each element type suffers from different variations, the final characteristics of the manufactured device are difficult to predict (10% variation from design value). So, the process variability is greatly reduced by utilizing capacitors alone (0.1% ratio mismatch). Second, a resistive feedback load reduces output impedance of amplifier, thereby reducing its equivalent open-loop gain. Utilizing capacitors for feedback solves this problem. Since DC operating points cannot be established through capacitors, a circuit switching mechanism determines periodic steady states of operation. For amplification of differential capacitance variations, either such a system must be used or a scheme based on modulation of a carrier applied to the differential capacitor.

This paper presents an inertial accelerometer device based on these technologies. Section I has briefly introduced MEMS base sensors. Section II briefly reviews similar devices. Section III outlines the top level modules of the accelerometer and Sections IV and V detail on the design of the mechanical sensing component and its interfacing charge amplifier, respectively. Section VI explains design methodologies, Section VII presents simulation results and Section VIII concludes the paper.

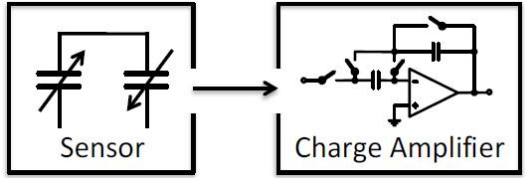

Fig 1. Inertial Accelerometer top level modules

## II. RELATED WORK

In [6] two variations of a SOIMUMPs based, single axis capacitive comb drive accelerometer are shown. Each is designed for a different thicknesses, 10μm and 25μm, for the structural layer. 3D models were developed for finite element simulations. Results were compared in terms sensitivity, capacitance and frequency response and displacement in the X, Y and Z axes.

A fully differential 0.25μ CMOS 14-bit ∑Δ digital interface for capacitive accelerometers was proposed in [7]. A low-noise switched-capacitor amplifier design, with correlated double sampling (CDS) readout circuit, is presented. A folded-cascode OTA was utilized as the core of each OpAmp in the switched capacitor charge amplifier and voltage integrator. The effectiveness CDS was evaluated for cancelation of low frequency flicker noise. The design was fabricated and integrated with a μG accelerometer sensor.

## III. OVERVIEW

The inertial accelerometer is broken down into the mechanical sensing element and the charge amplifier, as represented in Fig. 1. To design a digital output accelerometer, an analog to digital conversion would be required. Design for an ADC module is not presented in this paper.

To detect acceleration, the movable proof mass of the sensor, explained in Section IV, varies the value of a differential capacitor. A differential capacitor consists, generically, of two fixed plates and a movable plate in-between. Thus, this intermediate point causes capacitance variations relative to the fixed plates. The charge amplifier employs a switched capacitor (SC) topology that samples the charge variations of the differential capacitor, and translates this input into an amplified voltage output.

Project requirements dictated the following characteristics for the inertial accelerometer:
- Dynamic Range (DR): +/- 5G
- Bandwidth >800Hz
- Non-Linearity: <5%
- Resolution: <15mg/ √Hz
- Quality Factor (Q) = 0.5

The sensor itself plays the major role in determining dynamic range and bandwidth, as physical design is more restrictive than well studied amplifiers with high gain and frequency performance.

These two modules are detailed in the following sections, and their requirements and specifications explained.

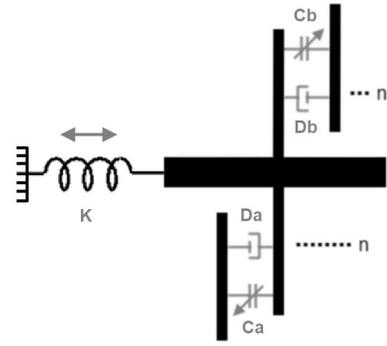

Fig 2. Lumped model of the mechanical behavior of the differential capacitance sensor. The moveable proof --mass holds 2 capacitor plates (arms)

## IV. MECHANICAL SENSOR

The sensor element consists of a movable proof mass supported by springs which allow movement, along a single axis, due to external acceleration. The proof mass itself holds movable capacitor plates, i.e. the intermediate plates between two fixed plates. This scheme of three parallel plates creates the equivalent of two variable capacitors in series. The movement of the proof mass causes a relative displacement between the fixed and movable parallel plates. Thus, a differential variation of capacitance occurs, where one capacitor increases while another decreases. The springs holding the proof mass impose a force-displacement behavior as dictated by Hooke's Law, and the surrounding air imposes the damping factor between the parallel plates. The lumped model is represented in Fig. 2, which also represents the capacitances that will vary in function of the mass displacement.

Thus, the sensor is a second order system, with a well-known mechanical response dictated by a spring constant, $k$, a damping constant, $b$ and a mass, $m$. The spring element introduces a force proportional to displacement, $F_{spring} = -x \times k$, and the damping introduces a reaction to velocity, given by $F_{damping} = v \times b$.

From this, the sensitivity, bandwidth and quality factor can be derived, and are given, respectively, by the following, where $m$ is the value of the proof mass, and $f$ the natural frequency:

$$S = m/k \qquad 2\pi f = \sqrt{k/m} \qquad Q = \sqrt{m \times k}/b \quad (1)$$

The spring and damping constants are, in turn, determined by the physical characteristics of the sensor, i.e., proof mass magnitude, spring dimensions and capacitor plate area. From this model, the minimum required displacement of the proof mass and spring constant were determined for the given DR and sensitivity while obeying Q and bandwidth.

The sensor has been designed targeting a thick silicon-on-insulator substrate based technology named SOIMUMPs (Silicon-On-Insulator Multi-User MEMS Process), commercially provided by CRONOS Inc. [11]. The process consists of several steps performed on a pre-prepared four layered Silicon-On-Insulator (SOI) wafer. The layers are, from top to bottom: 10μm or 25μm thick silicon, thin oxide (0.05 μm), substrate (400μm) and BOX. The silicon layer is used as the structural material. It is patterned and etched, via a Deep Reactive Ion Etching (DRIE) [12] process, to construct the mobile and fixed micromechanical structures. The minimum allowed dimension for patterned structures and structure spacing is 2μm.

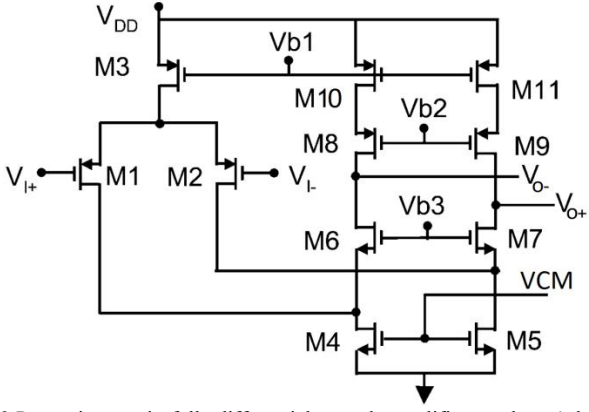

Fig.3 P-type input pair, fully differential cascode amplifier topology (adapted from [7])

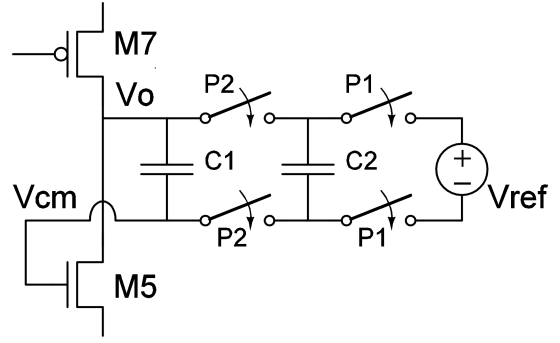

Fig.4 Dynamic Common Mode Feedback, for the right branch of the OpAmp, the same is repeated for the left branch.

For the sensor here presented, a 25μm thick SOI wafer was chosen, to maximize capacitor plate area and the respective differential capacitance variation. Section VI.A explains the design methodology of this module.

## V. READOUT CIRCUIT

The charge amplifier includes the Operational Transconductance Amplifier (OTA) itself and the switched capacitor sampling scheme. Since the utilized OTA is fully differential, it must also include Common Mode Feedback (CMFB), to set a stable DC operating point.

The target technology is AMS 0.35μm. This imposes a minimum of 0.35μm for the gate length and a maximum supply voltage of 3.3V.

### A. CMOS Fully Differential OTA

Fig. 3 represents the chosen topology for the OTA. It consists of a fully differential folded cascode with a P-type input pair. Specifications for the OTA are the typical required for an ideal amplifier: high gain, maximum excursion and low noise and offset.

The first stage gain should be as close as possible to unity to reduce input Miller effect. The second stage gain is provided by the common-gate FETs ($M_6$ and $M_7$). It can be shown that, for this circuit, the gain is independent of the biasing current. So, in function of the FET technology parameters only, the gain for one branch of the OTA is as follows (considering $A_1$ as a unitary gain, and where $R$ is the load into the source of $M_6$):

$$\frac{V_o^- - V_o^+}{V_i^+ - V_i^-} = A_1 \times A_2$$

$$A_1 \times A_2 = gm_1(r_{o1} \parallel r_{o4} \parallel R) \times gm_6 r_{o6} = \frac{2}{Vov_6 \times \lambda_6} \quad (2)$$

It is shown in Section VI.B how the channel modulation parameter, $\lambda$, and the threshold voltage, $Vt$, vary. To increase gain, $V_{ov}$ should be as small as possible, so, the bias voltages ($V_{b1}$ to $V_{CM}$) are set by the minimum $V_{ov}$ criterion. Design of adequate bandgap voltage references for biasing is out of the scope of this implementation. A small $V_{ov}$ is also required to achieve the maximum excursion, which implies larger FETs which also improves noise performance. The FETs that introduce the most noise are $M_4$, $M_5$, $M_2$ and $M_1$, as they are subject to the common-gate gain. Two sources of noise can be considered, thermal and flicker ($1/f$) noise.

### B. Common Mode Feedback

The common-mode (CM) feedback circuit is a dynamic capacitor scheme, shown in Fig 4. The two clock phases, P1 and P2, alternate the connections of capacitor C2 between a reference voltage, *Vref*, and the output and VCM nodes. Phase P1 imposes *Vref* on C2. Once the phase switches, C2 distributes charge to C1, maintaining a value of *Vref* plus $Vgs_{4,5}$, which, without CMFB, would change due to discharge of the parasitic output capacitors due to leakage (leading to two different DC operating points at each output node). For differential operation, charge can be shared between the two CMFB branches through the C1 capacitors, which share a common point at the gates of $M_{4,5}$. So, the CMFB has no effect on differential behavior.

For this circuit, C2 simply has to be large enough to negate parasitic capacitance effects, and C1 also serves as a pole compensation capacitance.

### C. Switched Capacitor Sampling

The front end circuit utilized is as shown in Figure 5. The topology is a fully differential SC circuit with Correlated Double Sampling. By using fully differential configuration, switch-induced errors and noise on the power rails are reduced, while doubling the input sensitivity and dynamic range.

The two $C_S$ capacitors represent the variable capacitance of the mechanical sensor. During clock phase P1, the two CDS capacitors, $C_{DS}$, eliminate input offset by charging to the same voltage. The $C_S$ capacitors are always charged to the same voltage also. Thus, their capacity variation will cause a variation of accumulated charge, which, during phase P2 will charge $C_A$ to different output voltages. This occurs because of CMFB which imposes the same common mode voltage at the input nodes, forcing the voltage difference to appear at the output.

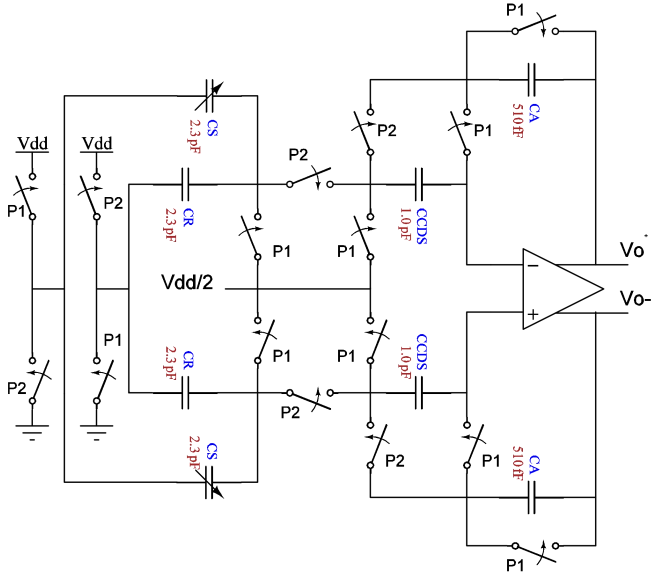

Fig.5 Dynamic Common Mode Feedback, for the right branch of the OpAmp. The same is repeated for the left branch.

By analyzing both clock phases, the differential output is given as:

$$(Vo^+ - Vo^-) = \frac{2V\Delta C}{C_A} + \frac{Vdd\,\Delta C}{C_A} \quad (3)$$

Where $V$ is given as:

$$V = \frac{Vdd}{2}\frac{(C_A + 3C_R - C_S)}{C_A + C_R + C_S} \quad (4)$$

So, it is shown that $C_A$ must be small in order to increase gain. Ideally, it should be equal to $\Delta C$ to saturate the output. However, the first term of equation 3, would negate the second, if $C_R$ was taken to be zero, since $C_A$ would be negligible relative to $C_S$. So, the $C_R$ capacitances are introduced to contradict this effect. If $C_R$ equals $C_S$, the gain is:

$$(Vo^+ - Vo^-) = \frac{2Vdd\,\Delta C}{C_A} \quad (5)$$

As with the voltage references, design and layout of the two non-overlapping clock trees, P1 and P2, is not presented here.

## VI. METHODOLOGY

### A. Mechanical Sensor

Designing the mechanical sensor entails determining dimensions for a number of features: the value of the capacitance and associated area and number of plates; the length and width of the springs supporting the proof mass, in function of the desired $k$, and the magnitude of the proof mass.

SOIMUMPs design rules guided the design process for the sensor, the most relevant being the minimal nominal distance allowed between silicon features, 2μm. This imposed the distance between capacitor plates and, consequently, total capacitor area required for a given capacity. The total plate separation is given by the sum of a fixed distance value, *gap*, plus the allowed movement range for the mobile plate, *da*. A compact design is preferred, so, plate separation should be as small as possible, to reduce needed area. Since design rules advised utilizing values above the minimum allowed for feature spacing, unless critically necessary, to avoid feature bridging, the minimum capacitor plate separation was set to be 2.5μm (including *gap* and *da*). Other restrictions include a minimum feature width of 6μm for released structures (to avoid length restrictions), and a minimum distance of 50μm from fixed structures to the trench etch boundary.

Since sensitivity and bandwidth are both determined by ratios of mass and spring constant, these 2 values correspond, actually, to only one degree of freedom. That is, neither mass nor $k$ can be freely varied to obey only either bandwidth or sensitivity individually. The same is true for the quality factor. Thus, there are only two degrees of freedom: the ratio of $k$ and $m$, and the value of the damping constant, $b$. Damping also determines the noise density (i.e. resolution), given by:

$$an = \sqrt{T \times k_b \times b}\,/(9.8 \times m) \quad (6)$$

By expressing sensitivity in terms of bandwidth we can determine the maximum allowed value for the sensitivity, to achieve a bandwidth of 800Hz (or higher), and vary it (which varies *da* accordingly to maintain DR constant) along with *gap*:

$$S_{\max} = 1/(2\pi f)^2 \quad (7)$$

In function of *da* and *gap* we can determine the required values for: the spring constant $k$, the required area for the capacitor and optimal mass value to obey all specification requirements. Fig 6 shows the minimal mass (μg), either the value required to obey noise requirements or the value that results from the needed capacitor area, and the optimal mass required to attain the wanted quality factor, for a capacitor value of 2pF. Capacitance is a given value for these curves, and was varied between 1pF and 4pF. Ideally, capacitance should be small to improve amplifier performance in terms of slew rate. The value of 2pF was chosen iteratively to accommodate viable

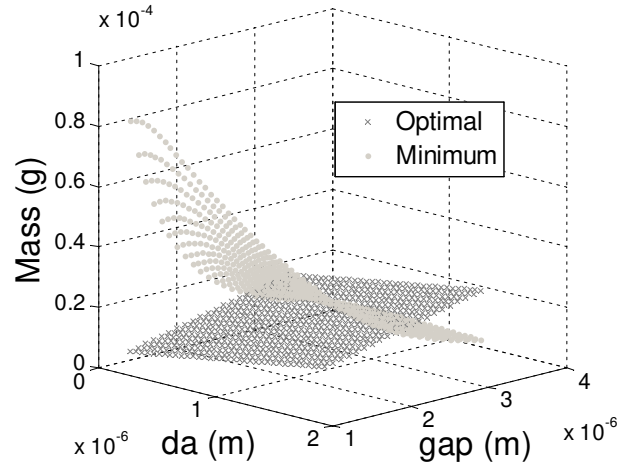

Fig 6. Required/optimal mass, in function of the fixed gap and the allowed movement range for the proof mass, for a capacitance of 2pF.

values for dimensions (that obeyed requirements).

Any combination of *da* and *gap* resulting in a minimal mass lower than an optimal mass is valid. However, the sum of the distances cannot be inferior to 2.5µm. Table I shows the variation of attained parameters by varying *da* (and *gap* accordingly). The fixed values of 2pF and 2.5 µm, fix the value of the needed capacitor area, 56.4µm$^2$, and its mass, 15.9µg. If the mass due to the capacitor exceeds the optimal mass, then the solution is invalid. So, 0.2µm for *da*, the first valid solution, was chosen. Characteristics are as show in the corresponding row of Table I. Quality factor and DR are implicitly obeyed. This allowed displacement corresponds to a variation of +/-155pF, for each capacitor, centered on the value of 2pF. No optimal solution was found when attempting to dimension the sensor with a silicon thickness of 10µm, as the required length and number of capacitor arms led to suboptimal combinations of capacitor mass and sensor dimensions.

Following this step, a generic architecture was established: symmetrical paired coiled springs holding several rows of movable capacitor arms. The spring constant, *k*, is determined by the length of the spring arm through an inverse cubic relationship. Since a single beam type springs require an unacceptable length to achieve values of the order shown in Table I, beams can be coiled to decrease *k*, maintaining length short.

To attain the needed values for *k* and *m*, the number of coils and length of the capacitor arms were varied. The total occupied area and the difference between total sensor length and width were analyzed in order to select the configuration resulting in an approximate square layout while utilizing the least area. All features were drawn with a minimum width of 6µm.

TABLE I
MECHANICAL SENSOR CHARACTERISTICS

| da | Cap. Mass | Opt. Mass | BW | Noise | Sens. | k |
|---|---|---|---|---|---|---|
| 0.1 | 15.9 | 13.2 | 2267 | 11.9 | 20.4 | 6.48 |
| 0.2 | 15.9 | 18.7 | 1603 | 8.46 | 40.8 | 4.58 |
| 0.3 | 15.9 | 22.9 | 1309 | 6.91 | 61.2 | 3.74 |
| 0.4 | 15.9 | 26.4 | 1133 | 5.99 | 81.6 | 3.24 |
| 0.5 | 15.9 | 29.6 | 1013 | 5.53 | 102 | 2.89 |

Characteristics of the mechanical system in function of the allowed plate movement, for a capacitor value of 2pF and a total plate distance of 2.5m.

*da* = µm; Cap. Mass, Opt. Mass = µg; BW = Hz; Noise = µG/√Hz, Sens. = nm/G, k = N/m

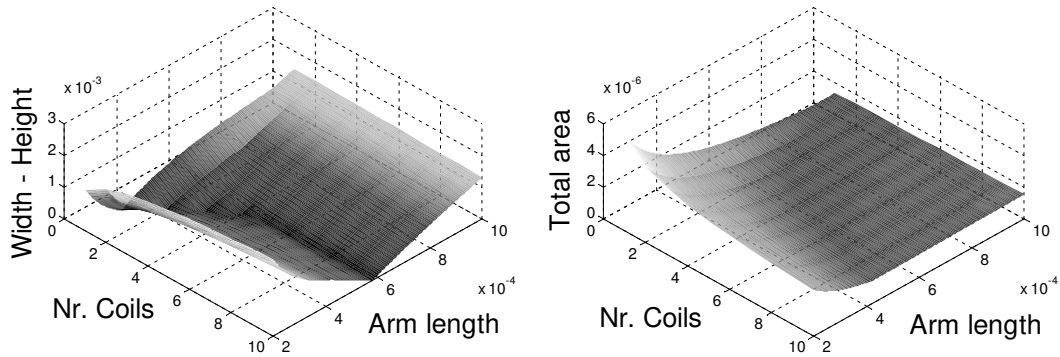

Fig 7. Required/optimal mass, in function of the gap and the allowed movable range for the movable arm of the sensor, for a capacitance of 2pF.

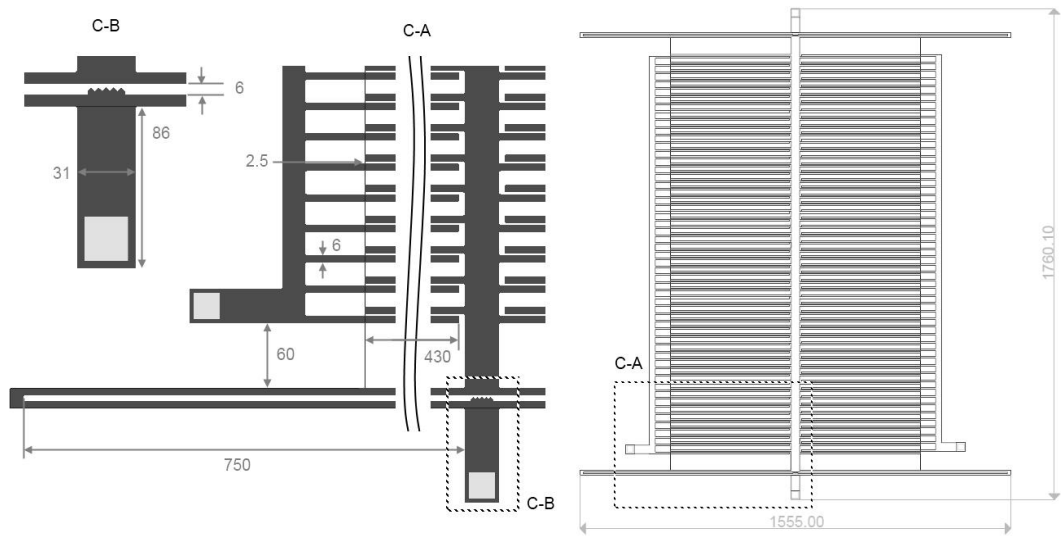

Fig 8. Mechanical design of the sensor. On the right, the entire structure is shown. C-A is a cut detail of the lower left corner and C-B the detail of the stopper structure. Dimensions given in µm.

The mobile plates create parasitic capacitors by pairing with the more distant complementary fixed plates. Each capacitor is thus given by the following, where $d_p$ is the distance from the mobile plate to the opposing fixed plate:

$$C_T = C + C_p = \frac{\varepsilon A}{d + \Delta d} + \frac{\varepsilon A}{d_p - \Delta d} \quad (8)$$

This distance must be such that non-linearity effects are reduced, as these parasitic capacitors will cause a reverse effect, negating capacity variations if they are of the same order of magnitude as the output capacitors. The charge amplifier will convert a capacitance difference to voltage, so, considering equation 8, this difference can be expressed as:

$$\Delta C = 2\varepsilon A \times \Delta d \frac{(d^2 - d_p^2)}{(d^2 - \Delta d^2)(d_p^2 - \Delta d^2)} \quad (9)$$

The linearity of this expression can be analyzed in function of $d_p$, for an input range of $\Delta d$ of -0.2μm to 0.2μm. It can be shown that for any value of $d_p$ above 2.5μm, non-linearity is below 0.5%. However, it can also be shown, from the derivative of (9), that the maximum capacitance difference stabilizes for values of $d_p$ above 10μm.

The minimums of the curves shown in Figure 7 are the regions that obey the design criteria. The optimal choice for architectural features is as follows: 2 coils for each spring, a spring beam length of 750μm (each with a constant of 2.29N/m and 0.26 μg), a total area of 22.4μm² (1449 μm by 1493 μm), capacitor arm length of 430μm, corresponding to 52 arms per capacitor, a proof mass of 31.2μm in width and length equal to the sensor length, providing the remaining 2.5μg of mass to achieve the optimal value of 18.7μg and a parasitic capacity separation of 13.5μm, which leads to 0.33% of non-linearity. Each capacitor has a true value of 2.37pF, due to the parasitic capacitors. The full structure is shown in Figure 8. C-B shows a stopper structure attached to the fixed anchor. This prevents the movable plates to come in contact with the fixed plates, due to an excessive external acceleration, which would cause a permanent bonding of both structures because of capillary and Van Der Waals forces [10].

### B. Readout Interface Circuit

In order to properly model the amplifier and CM detector, variability of transistor technology parameters were extracted from simulation, for both N and P type FETs. The evaluated parameters were the threshold voltage, $V_t$, the $K_n$ and $K_p$ parameters and the channel length modulation parameter, $\lambda$. Some of the extracted characteristics are show in Figure 9. To determine dimensions for the FETs, for a given current and $V_{gs}$, the extracted data was interpolated to compute the best values.

The bias current has no influence on gain, so a minimum value for the current is set by the necessary slew rate to respond adequately during the required sampling period. The required bandwidth is 800Hz. Adjusting this by a factor of 200 we arrive at a sampling frequency of 160 kHz. Multiplying by 10 to allow for a full charge or discharge of the feedback capacitors (0 to $V_{dd}$) within a fifth of a half of a clock period (5τ), we arrive at a minimum current of 12.5μA. The current was set to 80μA however, to account for parasitic capacity. Given this, the chosen minimum length for all FETs was 3.5μm, both to reduce noise and to maximize gain, due to reducing λ. This is also the point at which all characteristics stabilize. Following, the cascode branches were dimensioned to accommodate the chosen current and $V_{gs}$. The load impedances of $M_1$ and $M_2$ were computed and these 2 FETs were adjusted to attain a gain of 1. However, it was observed that, since $R$ was very large in comparison to $r_{o6}$, the second stage gain reduced to $gm_6 r_6$.

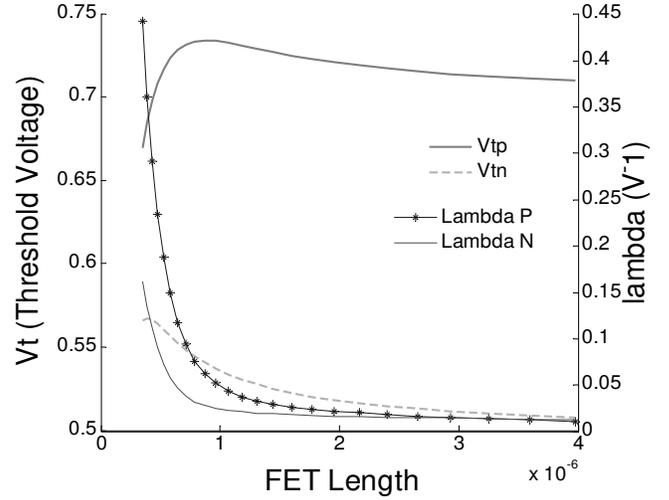

Fig 9. Threshold voltages and lambda parameters in function of FET Length

Thus, the maximum gain was estimated to be approximately 700. To increase gain, the input pair was re-dimensioned to increase first stage gain. Since the sensor is meant to operate at low frequencies, the Miller effect introduced by this is negligible. By increasing the size of $M_{1, 2}$, the noise contribution of these FETs decreases.

The design dimension of the FETs are slightly different from the computed values due to layout convenience, and are as follows: 294.3μm for $M_{10,11}$, 297μm for $M_{8,9}$, 85.8μm for $M_{6,7}$, 154.95μm for $M_{4,5}$, 274.4μm for $M_{1,2}$ and 457.8μm for $M_3$. Bias voltages, Vb1 to Vb3, are 2.41, 2.17 and 1.02. VCM should be established to 0.71V. $C_A$ is 510fF, $C_R$ equals Cs, i.e., 2.3pF, and $C_{DS}$ is set to 1pF, which is simply large enough above parasitic capacitance values. The switches are CMOS switches and simple N-type FETs when they are connected to ground. Switch transistor sizes were set so that their equivalent resistance coupled with the capacitances they were required to charge and discharge, were functional within the operating frequency of 160 kHz. The maximum allowed resistance, 156kΩ, resulted in 0.4μm and 2.5 μm for N-type and P-type FETs of the CMOS switches, respectively.

Figure 10 shows the layout for the OTA, including CMFB capacitors and switches. The total area (accounting for the substrate contact ring), is 30940μm² (238μm by 130 μm).

### VII. RESULTS

Simulations were run with both schematic and layout level circuits. Table II summarizes the observed characteristics for both simulations.

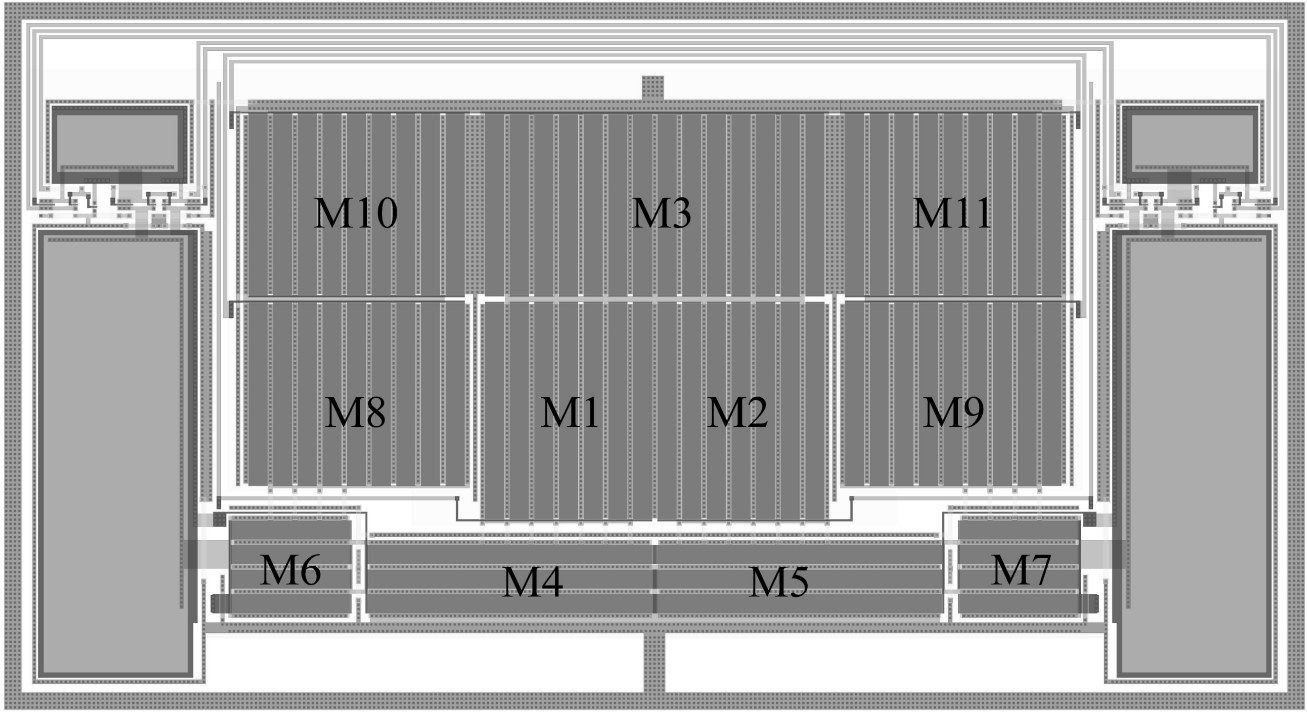

Fig 10. OTA layout, including CMFB and switches. The compensation capacitors could be reduced, as phase margin is sufficiently large.

To simulate acceleration input, the mechanical sensor was electronically emulated via an equivalent series RLC circuit. It can be shown that the sensor mass is equal to the inductance (18.47nH), the damping constant is equal to the resistor (585.8uΩ) and the inverse of the spring constant is the value of the capacity (0.218F). The resulting voltage across the capacitor equals input displacement. This voltage controlled the differential $C_S$ capacitors.

Figure 11 shows the bode plots for both simulations. A slight loss of gain was observed, which reduced the initial estimated phase and gain margins. Thus, the C1 compensation capacitors were adjusted, from 1.6pF to 1.83pF, to account for this after layout. The value of 1.6pF was the value observed to be sufficient for compensation at schematic level. The parasitic elements introduce a zero near the GHz range, which considerably lowers gain margin.

It was previous stated, from equation 3, that $C_A$ should equal ΔC in order to achieve maximum excursion. However, since the OTA saturates before *Vdd*, $C_A$ was set to 510fF, as shown in Figure 5, to limit differential output excursion to 2 volt (+/1 volt around the output common mode value of 1.64 volt), which is the point where the OTA output begins to distort.

The non-linearity estimated for the mechanical sensor (0.33%) improved due to a loss of gain near the excursion limits of the OTA, which negated the distortion effect.

It was observed that the parasitic capacitances introduced by the FETs themselves were considerable. The required slew rate to charge the output capacitors to the maximum excursion was of 3.2MV/s. Although the bias current was greatly over dimensioned to account for parasitic load, its effect still lowered achieved slew rate to only slightly above half the required value (4.5MV/s). Output CM is also affected, differing in a considerable 200mV from the expected value, due to a clock feed through effect, because the P1 switches have gate-to-drain capacity comparable to C2. So, C2 could be increased to correct for this effect.

Figure 12 displays the noise contribution curves for the entire system (sensor and OTA). Noise was analyzed up to half of the sampling frequency to account for aliasing. Due to aliasing, thermal noise from the OpAmp is the main source of noise. The sensor noise cuts off at frequencies above 1.583 kHz (from the RLC equivalent), which is the noise shaping observed in Figure 12. Due to this, the sensor introduces no noise aliasing with the used sampling frequency. The noise peak observed, at 160 kHz, is a replica of flicker noise due to noise aliasing introduced by the sampling.

TABLE II
PRE/POST-LAYOUT COMPARISON

| Parameter | Pre-Layout | Post-Layout |
|---|---|---|
| OTA Gain | 8.94k (79.03dB) | 8.59k (78.68dB) |
| OTA Gain Margin | -36.72 | -16.48 |
| OTA Phase Margin | 67.23 | 69.04 |
| Bandwidth | 5.00 | 4.90 |
| GPB | 44.72 | 42.07 |
| Non-Linearity | 0.183 | 0.077 |
| Total Sensitivity | 0.392 | 0.386 |
| Slew rate | 4.9 | 4.5 |
| Noise Density | 10.3 | 10.5 |
| Output CM | 1.83 | 1.64 |

Characteristics of the system, i.e., the switched capacitor circuit with the OTA, including CMFB and mechanical sensor input simulated via equivalent RLC.
Gain = V/V; Gain Margin = dB; Phase Margin = degrees; Bandwidth = kHz; GPB = MHz; Non-linearity = %; Total Sensitivity = V/G; Slew rate = V/s; Noise = µG/ √ Hz; Output CM = V

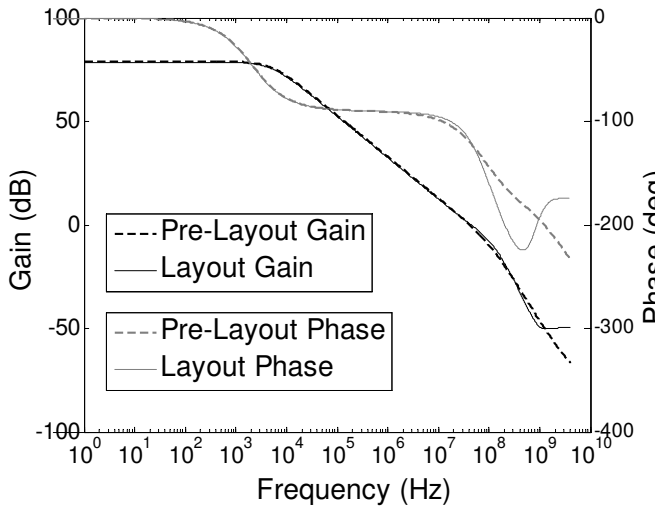

Fig 11. Pre and Post Layout Bode plot for the OTA

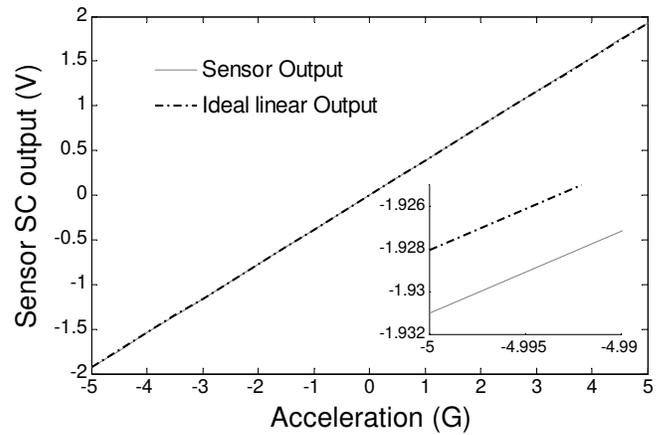

Fig 13. Pre and Post Layout Bode plot for the OTA

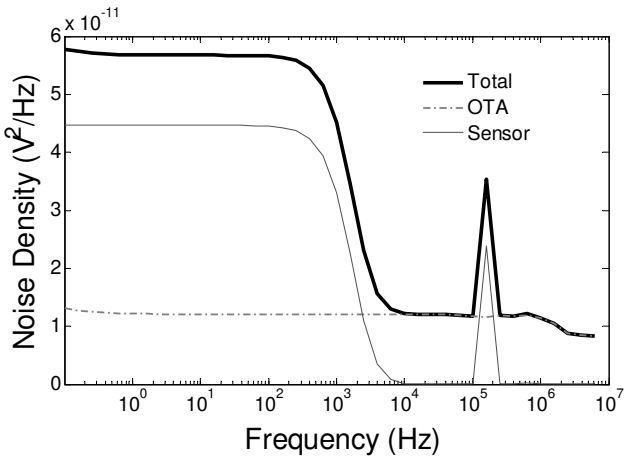

Fig 12. Pre-Layout Noise Density contributions and total noise of the accelerometer

The post-layout output characteristic of the system is shown in Figure 13. The curve shown was sampled at the output of the OTA at the end of the amplification clock phase (P2), with an ideal sample and hold. The dashed line represents the derivative at 0G. Shown is a detail of the nonlinearity at the maximum excursion.

## VIII. CONCLUSIONS

This paper presented the design, layout and layout simulation results for an inertial accelerometer, composed by a mechanical sensing element targeting a SOIMUMPs fabrication process and an SC charge amplifier. The designed OTA targets a 0.35μm CMOS technology.

The bandwidth of the system is determined by the sensor, which has a second order system behavior. Since the quality factor equals 0.5, there is no oscillation frequency. The OTA was the prominent noise source. The imposed requirements left very little design margin for the sensor. The mass due to the capacitor plates exceeded the proof mass required to obey the requirements for larger plate distance, but this distance could not be reduced bellow 2.5μm, which lead to a small design space.

Simulations indicate that the presented designs are within the required specifications.